# A working with a Silicon Position Sensitive Detector
# (dE%E calibration and isotopes separation)

V. Ostashko
*Institute for Nuclear Research, Kiev, Ukraine, ostashko@kinr.kiev.ua*

A. Tumino, S. Romano
*Dipartimento di Metodologie Chimiche e Fisiche per l'Ingegneria, Università di Catania, Catania, Italy*
*Istituto Nazionale di Fisica Nucleare, Laboratori Nazionali del Sud, Catania, Italy*

**Abstract**
A method for the calibration of a Silicon Position Sensitive Detector is described. From the combination of the signals collected at the ends of the resistive electrode, energy and angle of the hitting particle are reconstructed with a procedure which straightforwardly accounts for different amplifier chain gains and dc offset effects.

## 1. Introduction

Any reaction process to be studied requires a means by which the information about the involved particles can be carried out. A wealth of detectors has been developed with this aim, which allows to reconstruct at least the energy of the hitting particle.

Silicon Position Sensitive Detectors (SPSD) [1,2] provide also the information on the angle of the emitted particle, as known. A resistive layer acts as an electrode, either anode or cathode, which splits the energy signal in two parts. From the partition of this signal the emission angle of the hitting particle can be reconstructed. A correct event reconstruction requires an accurate knowledge of the energies and angles, thus a systematic calibration procedure is claimed. The two signals, collected from opposite ends of the SPSD resistive layer, are acquired through different pre-amp-amplifier-ADC combinations, which provide different amplification gains and dc offsets. The effects of these differences might be removed in preliminary steps, before combining the two signals for the calibration. The evaluation of the amplification gain requires a reference point in the detector to be taken. The signals at the two end-points can be feasible in order to calculate the two gains, but not always the end-points can be unambiguously determined because of the lack of detailed constructive information. A pulser is often used in order to empirically determine the offset, whose value can be deduced through a fit of the channel number versus the peak height. However the pulser itself, as well as any other device, can introduce some external offsets, thus increasing the number of undesired effects.

In this paper an original way of calibrating a SPSD is described. Starting from the combination of the two uncalibrated signals, the procedure accounts for all the mentioned effects during the calibration phase providing energy and angle of the detected particle in a straightforward way.

## 2. Schematic description of a Silicon Position Sensitive Detector

The calibration method, which will be described, was successfully applied to a SPSD [3] whose schematic picture is shown in Fig. 1.

The detector is made of a single wafer of doped n-type silicon. On the front face a p-type layer created by boron ion implantation acts as a resistive anode.

At both ends of the implanted anode, two readout contacts were created in order to deduce the position information from the partition of the energy signal. The cathode, on the back, consists of a $NN^+$ pseudo-junction, which acts as a non-injecting ohmic contact.

A reverse bias voltage, applied across the detector, depletes the bulk of the crystal. When an ion hits the detector, the energy release, due to the collision with bound electrons, causes a number of electron-hole pairs to be created in proportion to the deposited energy. The electric field within the silicon pulls electrons and holes apart, making them migrating to the front and back faces of the detector. A charge signal is therefore produced at both faces.

As already mentioned, the signal is taken from the front resistive layer, via contacts at each end of the strip, the wafer thickness being much smaller than its length.

The collected charge, which "sees" two paths to earth, divides into two fractions each one in inverse proportion to the distance between the hit position and the strip-end collecting such a fraction. The 1kΩ offset resistors ensure measurable signals to be collected wherever the ion hits the detector, otherwise a hit at one end of the strip would see a path of near zero resistance, almost all the charge following this route.

**3. Calibration procedure**

The charge created from an ion hit inside the detector is proportional to the deposited energy E, the amount of energy required to create an electron-hole pair being fixed. The proportionality between the energy deposit E and the charge created Q is expressed as:

$$E = Q/q \quad (1)$$

where q is the charge produced per MeV. The charge signal splits at the front strip, resulting into two strip-ends signals, $q_1$ and $q_2$, that, neglecting any charge loss, fulfil the following relation:

$$Q = q_1 + q_2 \quad (2)$$

If the total uniform resistance of a strip is $R_T$, and r represents the one at each end of the strip, then

$$q_1(X) = \frac{r + (0.5 + x)R_T}{2r + R_T} Q \quad (3)$$

$$q_2(X) = \frac{r + (0.5 - x)R_T}{2r + R_T} Q \quad (4)$$

*x* - being the distance of the hit point X from the strip centre (see Figure 1).

The digitalized information coming from ADCs, accessible during analysis data, is related to the original energy by the following relations

$$P_1(X) = \alpha \frac{r + (0.5 + x)R_T}{2r + R_T} E + P_{01} \quad (5)$$

$$P_2(X) = \beta \frac{r + (0.5 - x)R_T}{2r + R_T} E + P_{02} \quad (6)$$

α and β being the amplifier gains of the two pilot chains, and $P_{01}$ and $P_{02}$ the two dc offsets. Figure 2 shows an example of a two dimensional spectrum of uncalibrated variables $P_1$ versus $P_2$, obtained during a $^{12}C+^{12}C$ calibration run at 40 MeV of beam energy.

A grid with a number of slits of known geometry was placed in front of the detector in order to locate some reference regions. Several groupings of events along different kinematical loci can be recognized in the figure and used for the calibration. The most evident ones refer to the $^{12}C+^{12}C$ elastic and inelastic scattering, some others to the $^{12}C+^{12}C \to \alpha + ^{20}Ne^*$ reactions [4,5].

An initial knowledge of amplifier gains and dc offsets, introduced by the pre-amp-amplifier-ADC combination would allow performing the calibration procedure through the equations (5) and (6). As mentioned before, their evaluation is not always unambiguously achieved by means of standard procedures, which might introduce external effects. The present method, which is being introduced, is not affected at all by these effects, being based on general arguments which finally provide the required energy and position parameters.

Rearranging the expressions (5) and (6)

$$(P_1 - P_{01})(2r + R_T) = \alpha [r + (0.5 + x)R_T] E \quad (7)$$

$$(P_2 - P_{02})(2r + R_T) = \beta [r + (0.5 - x)R_T] E \quad (8),$$

and adding the equations (7) and (8)

$$(P_1 - P_{01} + P_2 - P_{02})(2r + R_T) =$$
$$[(\alpha+\beta)r + \alpha(0.5+x)R_T + \beta(0.5-x)R_T] E \quad (9)$$

a relation is carried out, showing the direct proportionality between the sum of the two acquired signals and the ion energy, for a given hit point. In order to obtain the information on the ion energy, the x value is obtained from the ratio:

$$\frac{P_1 - P_{01}}{P_2 - P_{02}} = \frac{\alpha [r + (0.5 + x)R_T]}{\beta [r + (0.5 - x)R_T]} \quad (10)$$

Making the following assignments

$$A_1 = -\frac{1}{R_T}(r + 0.5 R_T) ; \quad A_2 = \frac{\alpha}{\beta}\frac{1}{R_T}(r + 0.5 R_T)$$

$$A_3 = \frac{1}{R_T}(r - 0.5 R_T)(P_{01} - \frac{\alpha}{\beta} P_{02}); \quad A_4 = \frac{\alpha}{\beta}$$

$$(11)$$

$$A_5 = (P_{01} - \frac{\alpha}{\beta} P_{02}) \; ; \; A_6 = -\frac{P_{01} + P_{02}}{2r + R_T}$$

$$A_7 = \frac{(0.5\,\alpha + 0.5\,\beta) R_T}{2r + R_T} \; ; \; A_8 = \frac{(\alpha + \beta)}{2r + R_T}$$

from (9) and (10) the expressions providing the calibrated position and energy values are obtained

$$x = \frac{A_1 P_1 + A_2 P_2 + A_3}{P_1 + A_4 P_2 + A_5} \quad (12)$$

$$E = \frac{P_1 + P_2 + A_6}{A_7 + A_8 x} \quad (13)$$

Parameters $A_1 - A_8$ were automatically determined in the fitting program through the minimization of the $\chi^2$ value:

$$\chi^2 = \chi_x^2 + \chi_E^2 = \sum_i \left( \frac{X_i^{exp} - X_i^{calc}}{\Delta X_i^{exp}} \right)^2 + \sum_i \left( \frac{E_i^{exp} - E_i^{calc}}{\Delta E} \right)^2 \quad (14)$$

where $X_i^{calc}$ and $E_i^{calc}$ are given by eqs. (12) and (13) using $P_1$ and $P_2$ uncalibrated coordinates for each event;

$X_i^{exp}$ - is the linear distance of each slit of the grid from the center of the detector, determined from the geometry of the grid;

$E_i^{calc}$ - is the calculated energy of the detected particle associated with one of the loci identified in fig.2. Energy loss in both target and dead layers is accounted for;

$\Delta X_i^{exp}$ - represents the position resolution and is quoted as 0.5 mm. This number corresponds to half width of a slit and is comparable with the intrinsic resolution of a standard SPSD;

$\Delta E$ is the average accuracy in the energy determination. Its value, of the order of 50 - 100 keV, is chosen in such a way that the reduced $\chi_E^2$, i.e. $\chi_E^2$ divided by the number $N_D$ of degrees of freedom in the fitting procedure, is about 1. Since the relation $\chi_E^2 / N_D \sim 1$ is a classical quantitative determination for a good fit, we have chosen this value as a parameter of accuracy in energy determination.

The $\chi^2$ minimization procedure has also proven to be feasible for a precise determination of the thickness of dead layers crossed by the particles before being stopped inside the active region of the detector. The thickness of the dead layers is varied until the best $\chi^2$ is obtained. Further technical cross checks on the detectors have validated the capabilities of this procedure.

It is worth stressing that there is no need to preliminarily deduce amplifier gains and dc offsets, since they are implicitly accounted for in the global calibration procedure. The position of each particle hitting the detector has to be converted into an angle relative to the beam direction. The following expression allows for determining the angle:

$$\theta = \arctan \frac{x}{d_c} + \theta_c \quad (15)$$

$d_c$ and $\theta_c$ being the distance from the target and the central angle of the detector, respectively.

Figure 3 shows the events reported in figure 2, scattered on the two-dimensional plane of the calibrated variables, angle and energy. The horizontal alignment of the angular regions located by the slits of the grid makes sure about the independence of the angular variable on the energy of the detected ion.

If the angle capture of the *dExE*-telescope is big, and with PSD-detector too, there is present complex problem to separate of different isotopes events for same elements. It was connect with it, that particles of same isotope and exit different angle are crossing different thickness of E-detector really. It is brought to straggling of dE-spectrum and the boundary of isotopes separating have angle depending of particle exit from target on the dE vs E plane.

For calibration we use reactions with simple isotopes set in exit channel and big cross-section and situation is clear. In investigation experiments we have not information about isotope picture and cross-section is not big too. In this case we have check simple method for separation of different isotopes of certain element.

We are using code STRAGG7 [6] for calculations of energy losses. This program allows calculation of energy losses for complex system. In formal it means that we can calculate energy losses for beam particle with mass isn't whole number. In practically, for energy value in E-detector (PSD) we are selecting (fitting) the value of mass of particle from condition that calculation energy losses of this particle with this mass and with its real way length in dE-detector (PSD-detector gives this possibility) is the same to registering value in dE-detector.

Figure 4a shows the example of projection on dE-axis of the part of two-dimensional dE%E spectrum for carbon region and energy in E-detector near 70 MeV and figure 4b shows the histogram of mass of those events (isotopes of carbon) calculating with method are described above.

### 4. Conclusions

An original method to calibrate a Silicon Position Sensitive Detector was described. Position and energy information for the hitting ion are reconstructed from the analysis of the signals collected at the ends of the resistive electrode. This method provides a general calibration procedure, which does not need preliminary information of the amplifier gains of the two chains as well as of the dc offsets. The expressions for the calibrated energy and position values are finally given in terms of the acquired signal heights and several parameters whose best values are simply obtained by means of a multiple regression treatment.

### Acknowledgments

One of the authors (V.V.O.) is grateful to L.N.S. for hospitality and financial support during the experiments.

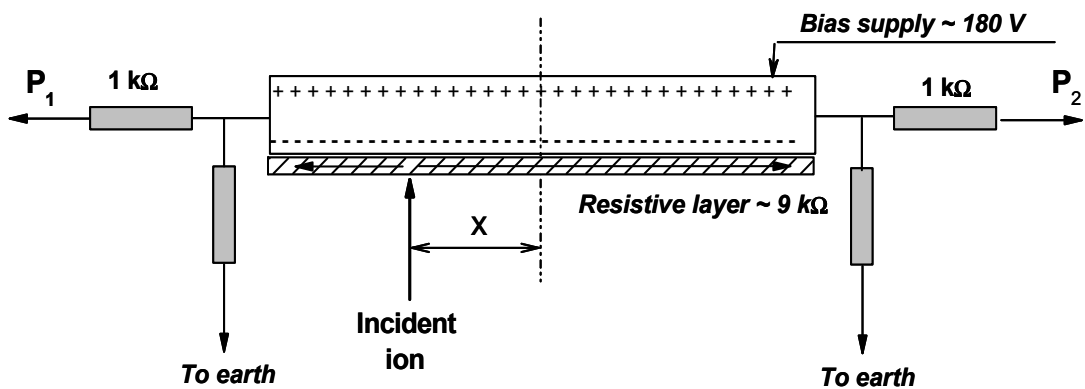

**Figure 1:** *Schematic picture of a SPSD detector with its electrical connections*

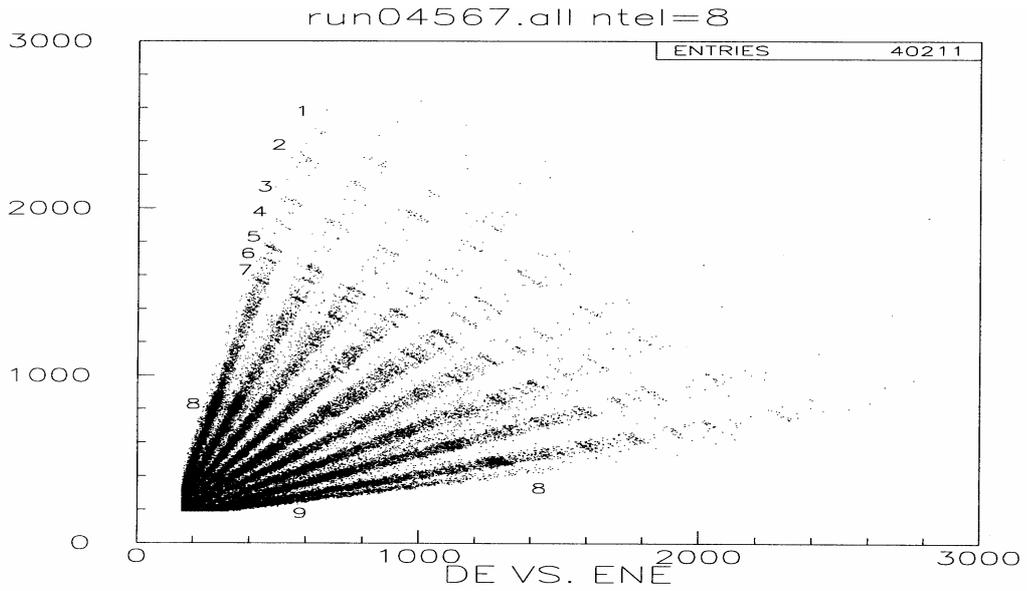

**Figure 2.** *Two dimensional spectrum of uncalibrated variables $P_1$ versus $P_2$. Loci from 1 to 7 refer to the $^{12}C$ ($^{12}C$, alpha)$^{20}Ne^*$ reactions, feeding several states of $^{20}Ne$; loci 8 and 9 are associated with the $^{12}C$ ($^{12}C$,$^{12}C$)$^{12}C$ and H($^{12}C$,H)$^{12}C$ reactions respectively*

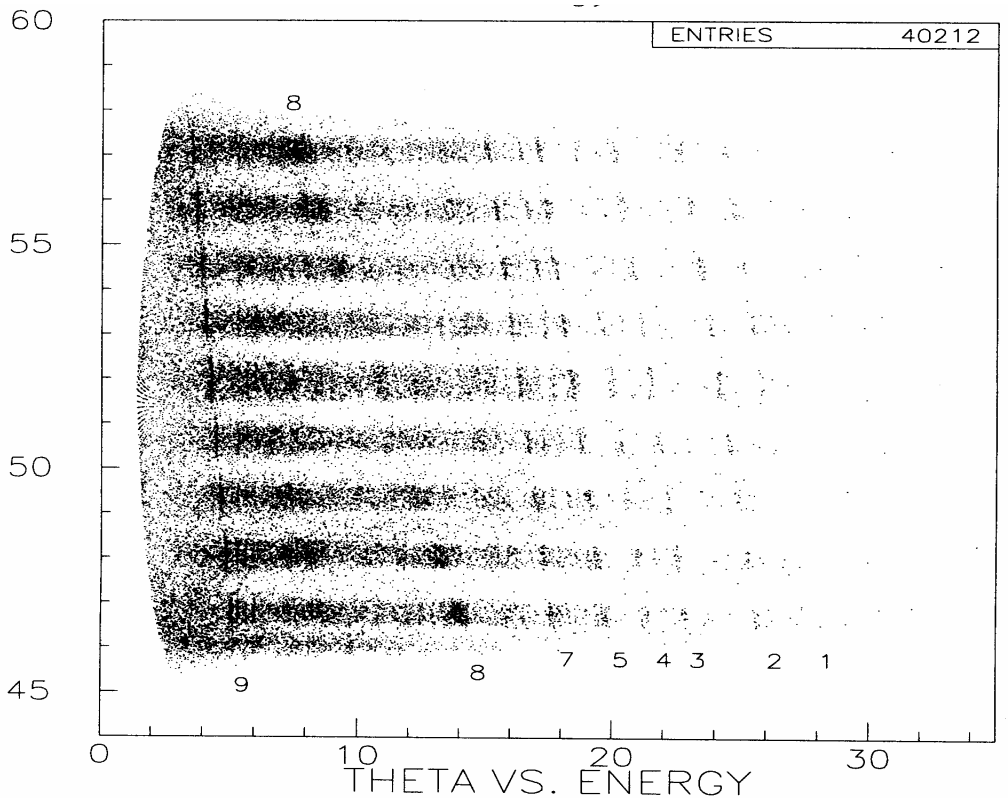

**Figure 3:** *Two-dimensional spectrum of the calibrated variables, angular position $\theta$ versus energy E. The numbers are associated with experimental loci as explained in the caption of Figure 2.*

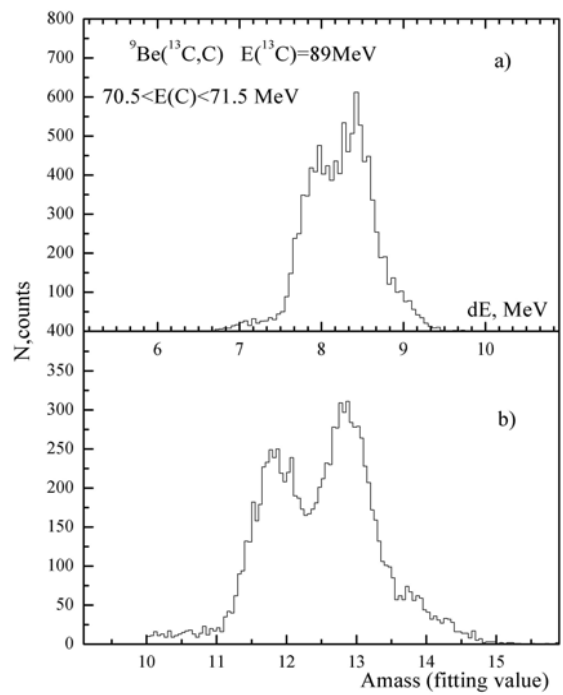

**Figure 4: a)** projection on dE-axis of the part of two-dimensional dE%E spectrum for carbon region and energy in E-detector near 70 MeV
  **b)** Histogram of calculating mass for those events (look the text).